# Decay of Rabi oscillations induced by magnetic dipole interactions in diluted paramagnetic solids

E. I. Baibekov

*Kazan Federal University, 420008 Kazan, Russian Federation*

**Abstract**

Decay of Rabi oscillations of equivalent spins diluted in diamagnetic solid matrix and coupled by magnetic dipole interactions is studied. It is shown that these interactions result in random shifts of spin transient nutation frequencies and thus lead to the decay of the transient signal. Averaging over random spatial distribution of spins within the solid and over their spectral positions within magnetic resonance line, we obtain analytical expressions for the decay of Rabi oscillations. The rate of the decay in the case when the half-width of magnetic resonance line exceeds Rabi frequency is found to depend on the intensity of resonant microwave field and on the spin concentration. The results are compared with the literature data for $E'_1$ centers in glassy silica and $[AlO_4]^0$ centers in quartz.

## I. INTRODUCTION

Rabi oscillations (or transient nutations) represent transitions between the states of quantum dipole driven by electromagnetic wave of resonant frequency [1]. Early observations of Rabi oscillations (RO) were accomplished on nuclear and electronic magnetic dipoles in NMR [2] and ESR [3] respectively, and on electric dipole moment of an atom in optical resonance [4]. The successful observation of RO imposes a number of severe constraints on experimental apparatus: low response time and high sensitivity of radiofrequency circuits, narrow bandwidth and high intensity of the generated electromagnetic field [2]. With the development of magnetic resonance techniques (pulsed NMR and ESR, spin echo) it became possible to observe RO at higher power levels. High-resolution RO detection with the transient pulse increment of the order of a few ns is now a routine [5]. Thus, it is now possible to observe RO in a variety of systems, including those with coherence times < 1 μs, such as electronic spins in molecular magnets [6]. In magnetic resonance, a comparison of the decay time $\tau_R$ of RO and phase relaxation time $T_2$ reveals how transient nutations driven by resonant microwave (MW) field affect coherence dynamics of the system of spins. Apart from purely academic interest, the answer to this question has an important practical application. In quantum computation, several perspective implementations of a quantum bit (qubit) include electronic or nuclear spin qubits [7] and trapped ion-based optical qubits [8]. They are operable with the application of electromagnetic transient pulses. During a short time interval of the pulse a qubit is in transient regime, i.e. accomplishes RO. If the number of successive operations on the qubit is sufficiently high, one must take into account not only the decoherence induced during time intervals between the pulses, but the one accumulated during RO.

In magnetic resonance, the observed decay of RO is known to arise from various reasons: (i) static $B_0$ [2] and microwave $B_1$ [2,9,10] field inhomogeneities, (ii) radiation damping [11], (iii) inhomogeneous broadening of magnetic resonance line [2], (iv) spin-spin and spin-lattice interactions [2]. However, (i) and (iii) lead only to the distribution of Rabi frequencies of spins in the solid. Of the above-mentioned mechanisms only (iv) result in "true" decoherence, i.e. irreversible change of a single spin state due to its interaction with the environment. Typically, at least at liquid helium temperatures, spin-lattice relaxation times $T_1$ exceed phase relaxation times $T_2$. Consequently, spin-spin interactions are of primary interest in the study of decoherence during the transient regime. At sufficiently low spin concentrations, the most important are long-range magnetic dipole (MD) interactions. Further, we can distinguish between the spins of *A* and *B* types [12,13]. The first, *A*, have Larmor precession frequencies close to resonance and thus are driven by the resonant transient pulse. The second group, *B*, contains spins with precession frequencies far from resonance. RO decay of a single *A* spin interacting with the bath of *B* spins was considered in [14]. An example of such a system is a single electronic spin coupled to a bath of nuclear spins. Completely different dynamics has the system of only *A* spins. There the transient signal is recorded not only from a single spin but from a great number of equivalent spins interacting with each other. While in [14] internal bath dynamics was represented via random field with phenomenological correlation function, we present here a detailed *ab initio* calculation of dynamics of *A* spin bath in transient regime. We will show that the decoherence in *A* spin bath, under appropriate conditions, is enhanced by the external resonant MW field, in contrast with the *B* spin bath where the corresponding RO decay rate was found to decrease with the increase of MW field intensity [14].

## II. THEORY

Let us consider a diamagnetic solid matrix containing a small concentration of magnetic impurity (e.g. paramagnetic ions, crystal defects or nuclear spins). In the common case when the ground state of a magnetic center is an isolated doublet, one introduces (pseudo-)spin $S = 1/2$. We accept also that the interaction of $j$-th center in the solid with the external magnetic field is represented via almost isotropic $g$-tensor: $g^j_{\alpha\beta} \approx g_e \delta_{\alpha\beta}$. All the subsequent calculations can be easily generalized for highly anisotropic $g$-tensor. In the presence of static $\boldsymbol{B}_0 \parallel z$ and resonant transverse microwave $\boldsymbol{B}_1 \cos\omega_0 t \parallel x$ magnetic fields ($B_0 \gg B_1$), the Hamiltonian of a system of spins coupled by MD interactions is as follows:

$$H = H_0 + H_1 + V = \sum_j \left(H_0^j + H_1^j\right) + \frac{1}{2}\sum_{j\neq k} V^{jk}, \tag{1}$$

where $H_0^j = g^j_{zz}\mu_B B_0 S_z^j \equiv \hbar\omega_j S_z^j$, $H_1^j = g_e\mu_B B_1 S_x^j \cos\omega_0 t \equiv 2\hbar\Omega_R \cos\omega_0 t$, $\mu_B$ is Bohr magneton. Because of different microscopic environments at the vicinities of spin positions, spatial inhomogeneity of $B_0$, dipolar interactions with magnetic impurities of other species inside the sample, etc. [11,15], Larmor angular precession frequencies of spins $\omega_j$ are distributed around $\omega_0$ in magnetic resonance line of half-width $\sigma$. Especially for electronic spins in ESR, $\sigma$ can well exceed Rabi angular frequency $\Omega_R$. MD interaction between $j$-th and $k$-th spins connected by radius-vector $\boldsymbol{r}_{jk}(x_{jk}, y_{jk}, z_{jk})$ equals:

$$V^{jk} = \frac{g_e^2 \mu_B^2}{r_{jk}^3}\left\{\left(\boldsymbol{S}^j \cdot \boldsymbol{S}^k\right) - \frac{3\left(\boldsymbol{S}^j \cdot \boldsymbol{r}_{jk}\right)\left(\boldsymbol{S}^k \cdot \boldsymbol{r}_{jk}\right)}{r_{jk}^2}\right\} \equiv \hbar \sum_{\alpha,\beta=x,y,z} A^{jk}_{\alpha\beta} S_\alpha^j S_\beta^k. \tag{2}$$

At the moment $t = 0$ when the oscillating resonant field $B_1 \cos\omega_0 t$ is turned on, the spin system is described by the initial density matrix of the form

$$\rho(0) \approx \frac{1}{2^N}\prod_j \left(1 - \frac{\hbar\omega_0 S_z^j}{k_B T}\right), \tag{3}$$

where $N$ is the number of spins, $T$ is the temperature of thermostat, $k_B$ is Boltzman constant, and $\hbar\omega_0 \ll k_B T$.

A transformation of Eq. (1) into the reference frame $RF$ rotating around $z$ axis of laboratory reference frame $LF$ with the angular frequency $\omega_0$ yields

$$H' \simeq \hbar\sum_j \left(\delta\omega_j S_z^j + \Omega_R S_x^j\right) + \frac{\hbar}{2}\sum_{j\neq k} A^{jk}_{zz}\left(S_z^j S_z^k - \frac{1}{2}S_x^j S_x^k - \frac{1}{2}S_y^j S_y^k\right), \tag{4}$$

where $\delta\omega_j = \omega_j - \omega_0$ is the shift of Larmor frequency of $j$-th spin from resonance, and we have omitted oscillating time-dependent terms in $H'$. Let us now introduce a new set of local axes $\tilde{x}^j, \tilde{y}^j, \tilde{z}^j$ (see Fig. 1), such that:

$$\tilde{S}_x^j = \left(\delta\omega_j S_z^j + \Omega_R S_x^j\right)/\Omega_j \equiv \alpha_j S_z^j + \beta_j S_x^j, \quad \tilde{S}_y^j = S_y^j, \quad \tilde{S}_z^j = \beta_j S_z^j - \alpha_j S_x^j. \tag{5}$$

The Hamiltonian (4) written in new spin projections is

$$H' = \hbar \sum_j \Omega_j \tilde{S}_x^j + \frac{\hbar}{2} \sum_{\substack{j \neq k \\ \alpha\beta}} \tilde{A}_{\alpha\beta}^{jk} \tilde{S}_\alpha^j \tilde{S}_\beta^k, \tag{6}$$

where $\Omega_j = \sqrt{\delta\omega_j^2 + \Omega_R^2}$ is the nutation frequency of $j$-th spin detuned by $\delta\omega_j$ from resonance. In the case of sufficiently small spin concentrations, we leave only the terms of MD interactions $\sim \tilde{S}_x^j \tilde{S}_x^k$ in Eq. (6) that are secular to the interaction with the external magnetic field in $RF$. As we will see, these terms result in random shifts of spin nutation frequencies. We neglect nonsecular terms responsible for mutual spin flip-flops that are significant only when the average local field at the spin site induced by nearby spins is comparable with $\sigma$.

The response of a spin system to the driving resonant MW field is given by time evolution of its magnetic moment:

$$\boldsymbol{M}(t) = g_e \mu_B \text{Tr}\left\{e^{-iH't/\hbar} \rho(0) e^{iH't/\hbar} \sum_j \boldsymbol{S}^j\right\}. \tag{7}$$

It is easier to calculate the trace in (7) in the basis of $\tilde{S}_x^j$ eigenfunctions $|\pm 1\rangle = \left(|\uparrow\rangle \pm |\downarrow\rangle\right)/\sqrt{2}$ rather than $\tilde{S}_z^j$ eigenfunctions $|\uparrow\rangle, |\downarrow\rangle$. For instance, $y$-projection $M_y(t)$ equals:

$$M_y(t) = \frac{g_e \mu_B}{2^N} \sum_{\{m_l\}} \prod_l \langle m_l | \left\{ \prod_{j'} \left(1 - \frac{\hbar\omega_0}{k_B T}\left(\beta_{j'}\tilde{S}_z^{j'} + \alpha_{j'}\tilde{S}_x^{j'}\right)\right) e^{iH't/\hbar} \sum_j \tilde{S}_y^j e^{-iH't/\hbar} \right\} \prod_l |m_l\rangle, \tag{8}$$

where $\prod_l |m_l\rangle$ is the $N$-spin state described by a set $\{m_l\} = (m_1, m_2, \ldots, m_N)$ of spin projections, $m_l = \pm 1$, and the summation is performed over all possible sets $\{m_l\}$. The terms linear in $\alpha_{j'} \sim \delta\omega_{j'}$ disappear after the average over the spectral position $\omega_{j'}$ of $j'$-th spin in the symmetric resonance line centered at $\omega_0$. Calculation of matrix elements in Eq. (8) yields

$$M_y(t) = -\frac{g_e \mu_B \hbar \omega_0}{2^{N+1} k_B T} \sum_j \beta_j \sum_{\{m_l\}'} \sin\left[\left(\Omega_j + \frac{1}{2}\sum_{k(\neq j)} \tilde{A}_{xx}^{jk} m_k\right) t\right], \tag{9}$$

where $\{m_l\}' = (m_1, \ldots, m_{j-1}, m_{j+1}, \ldots, m_N)$ denotes the configuration of $N-1$ spins. The obtained expression is intuitively obvious: $j$-th spin oscillates around $\tilde{x}^j$ axis in $RF$ with angular frequency $\Omega_j + \frac{1}{2}\sum_{k(\neq j)} \tilde{A}_{xx}^{jk} m_k$. The shift $\tilde{A}_{xx}^{jk} m_k / 2$ of its nutation frequency originates from the local field at its site induced by $k$-th spin in state $|m_k\rangle$. Performing summation over all possible spin configurations $\{m_l\}'$, we obtain:

$$M_y(t) = \frac{M_0}{N} \sum_j \beta_j \sin\Omega_j t \prod_{k(\neq j)} \cos(\tilde{A}_{xx}^{jk} t/2), \qquad (10)$$

where $M_0 = -\dfrac{N g_e \mu_B \hbar \omega_0}{4 k_B T}$ is the initial magnetization of the spin system.

The factor $\prod_{k(\neq j)} \cos(\tilde{A}_{xx}^{jk} t/2)$ in Eq. (10) is responsible for the damping of RO because of MD interactions. Further calculations require averaging over spatial distribution of spins within the solid and over their spectral positions within the resonance line. We assume that (i) spins occupy random positions $r_k$ within the volume $V$, (ii) their frequencies $\omega_k$ are distributed within the symmetric resonance line of spectral density $g(\omega_k)$ centered at $\omega_0$, (iii) $r_k$ and $\omega_k$ do not correlate. The averaging over $r_k$ and $\omega_k$ in the limit $N, V \to \infty$ yields:

$$\prod_{k(\neq j)} \left\langle \cos(\tilde{A}_{xx}^{jk} t/2) \right\rangle_{r_k, \omega_k} = \left\{ 1 - \frac{C}{N} \int d\omega_k g(\omega_k) \int_\infty d^3 r_k \left[1 - \cos(\tilde{A}_{xx}^{jk} t/2)\right] \right\}^{N-1} =$$
$$= \exp\left\{ -C \int d\omega_k g(\omega_k) \int_\infty d^3 r_k \left[1 - \cos(\tilde{A}_{xx}^{jk} t/2)\right] \right\}, \qquad (11)$$

where $C = N/V$ is the spin concentration. Substituting $\tilde{A}_{xx}^{jk}$ as a function of $r_{jk}$ and performing the integration over $r_k$ in Eq. (11), one obtains

$$M_y(t) = M_0 \int d\omega_j g(\omega_j) \beta_j \sin\Omega_j t \exp\left(-\Delta\omega_d t \int d\omega_k g(\omega_k) |\alpha_j \alpha_k - \beta_j \beta_k / 2|\right), \qquad (12)$$

where $\Delta\omega_d = \dfrac{4\pi^2 g_e^2 \mu_B^2 C}{9\sqrt{3}\hbar}$ is known as the static half-width of the resonance line due to MD interactions [13]. The above statement that non-secular parts of MD interactions are negligible as compared with secular ones corresponds to the condition $\Delta\omega_d \ll \sigma$.

Given the line shape $g(\omega)$, one can calculate numerically the average magnetization (12) for the given values of $C$, $\Omega_R$ and $\sigma$. However, it is possible to obtain analytical expressions for $M_y(t)$ in two practically important approximations:

a) The case of weakly inhomogeneously broadened resonance line (or high MW field): $\sigma \ll \Omega_R$. All spins in the line nutate with the same frequency $\Omega_R$, and

$$M_y(t) = M_0 e^{-\Delta\omega_d t/2} \sin\Omega_R t. \qquad (13)$$

Thus, when the resonance line is fully excited, MD interactions cause exponential decay of RO with the rate $\tau_R^{-1} = \Delta\omega_d/2$.

b) Strongly inhomogeneously broadened resonance line (weak MW field): $\sigma \gg \Omega_R$. The resonance line is partially excited. Roughly, only the spins with Larmor frequencies falling into the range $[\omega_0 - \Omega_R, \omega_0 + \Omega_R]$ are involved in RO. At sufficiently small spin concentrations, the condition $\Omega_R \gg \Delta\omega_d$ is also valid. When $\sigma t \gg 1$, i.e. almost immediately after the transient regime is turned on,

$$M_y(t) = \pi M_0 g(\omega_0) \Omega_R e^{-\beta_d \Omega_R t} J_0(\Omega_R t), \tag{14}$$

where $J_0(z)$ is Bessel function of the 1$^{st}$ kind. It is remarkable that the parameter $\beta_d$ has the same form for both Gaussian $g_G(\omega) = (2\pi\sigma^2)^{-1/2} \exp[-\delta\omega^2/2\sigma^2]$ and Lorentzian $g_L(\omega) = \sigma/[\pi(\delta\omega^2 + \sigma^2)]$ distributions:

$$\beta_d = \Delta\omega_d g(\omega_0) \ln\frac{2\sigma}{\Omega_R}. \tag{15}$$

Results of numeric calculations of the decay rate obtained from Eq. (12) (assuming Gaussian line shape) for arbitrary values of $\Omega_R/\sigma$, together with the asymptotic solutions for $\sigma \ll \Omega_R$ and $\sigma \gg \Omega_R$, are presented in Fig. 2. The time dependence of transverse magnetization calculated from Eq. (14) for $C = 10^{17} \text{cm}^{-3}$, $\sigma = 2\pi \cdot 1\,\text{MHz}$ and $\Omega_R = 2\pi \cdot 100\,\text{kHz}$ is shown in Fig. 3.

RO are detected in experiment either through transverse $(M_\perp(t))^2 = (M_x(t))^2 + (M_y(t))^2$ (e.g. two-photon resonance technique [16,17]) or longitudinal $M_z(t)$ (e.g. transient pulse followed by spin echo sequence [18]) magnetization. Following the same argumentation, one obtains $M_x(t) = 0$, in agreement with Bloch model predictions [2], and thus $M_\perp(t) = M_y(t)$. Similarly,

$$M_z(t) = \begin{cases} M_0 e^{-\Delta\omega_d t/2} \cos\Omega_R t, & \Delta\omega_d \ll \sigma \ll \Omega_R, \\ \pi M_0 g(\omega_0) \Omega_R e^{-\beta_d \Omega_R t} j_0(\Omega_R t), & \Delta\omega_d \ll \Omega_R \ll \sigma, \end{cases} \tag{16}$$

where $j_0(z) = \int_z^\infty J_0(z) dz$.

Particularly interesting is the case of strong inhomogeneity $\Delta\omega_d \ll \Omega_R \ll \sigma$ common in magnetic resonance experiments. There, after a number of oscillations, i.e. $\Omega_R t \gg 1$, one can use asymptotic representation $j_0(z \gg 1) \simeq J_0(z + \pi/2) \simeq \sqrt{2/\pi z} \cos(z + \pi/4)$. In this case, there appear two contributions into RO decay:

(i) Polynomial damping $\sim (\Omega_R t)^{-1/2}$ originating from the superposition of transient nutations of spins with different Larmor frequencies; this contribution is absent in the case of weak inhomogeneity $\sigma \ll \Omega_R$.

(ii) Exponential damping $\sim e^{-\beta_d \Omega_R t}$ caused by MD interactions between the spins driven by transient pulse; since $\beta_d$ depends on $\Omega_R$ through slowly-varying logarithmic function, the damping rate $\tau_R^{-1} = \beta_d \Omega_R$ is roughly linear in Rabi frequency in the specific range of $\Omega_R$. When $\sigma \ll \Omega_R$, the damping rate is independent of $\Omega_R$, cf. Eq. (13).

The contribution (i) is well-known in the field of magnetic resonance (see, e.g., [2]) and does not lead to decoherence of spin states; (ii) reflects decoherence inside a bath of equivalent spins coherently driven by resonant MW field and coupled by MD interactions.

The above results can be interpreted semi-quantitatively as follows. Firstly, we consider the case $\sigma \ll \Omega_R$ when all spins have nearly the same nutation frequency $\Omega_R$. A given spin $k$ oscillates around $x$ axis in $RF$, where its time-averaged spin projection $\langle S_x^k \rangle$ is nonzero. Thus, via the terms $\sim S_x^j S_x^k$ in MD interaction, all $N-1$ spins induce a static random local field at the site of $j$-th spin. Just as the term $\sim S_z^j S_z^k$ in Eq. (4) would lead to the rms shift of Larmor frequency of $j$-th spin equal to $\Delta\omega_d$ [13], the term $\sim S_x^j S_x^k / 2$ results in rms shift $\Delta\omega_d / 2$ of its nutation frequency, and finally to the decay rate $\tau_R^{-1} = \Delta\omega_d / 2$ of RO. When $\sigma \gg \Omega_R$, we must reduce the number of spins involved in transient nutations from $N$ to $\simeq N\Omega_R/\sigma \simeq Ng(\omega_0)\Omega_R$, and the decay rate of RO is now estimated as $\tau_R^{-1} \sim \Delta\omega_d g(\omega_0) \Omega_R$, which is comparable to the one obtained in Eqs. (14), (15).

### III. DISCUSSION

It is noteworthy to compare our results with the existing models of RO decay. The first description of RO decay was presented in the framework of Bloch model [2]. As above, we admit that relaxation in the spin system is caused by interspin MD interactions, that $T_1 \gg T_2$, and consider again two cases of weak ($\sigma \ll \Omega_R$) and strong ($\sigma \gg \Omega_R$) inhomogeneity. Then, according to Bloch equations, the transverse magnetization in transient regime exhibits the damped oscillations of the form [2]:

$$M_\perp(t) \sim \begin{cases} e^{-t/2T_2}\sin(\Omega_R t), & \sigma \ll \Omega_R, \\ e^{-t/2T_2}J_0(\Omega_R t), & \sigma \gg \Omega_R, \end{cases} \quad (17)$$

with the damping rate $\tau_R^{-1} = T_2^{-1}/2$ equal in both cases and independent of $\Omega_R$, in contrary to Eqs. (13), (14). This misleading result is due to simplicity of Bloch equations that incorporate relaxation processes via the phenomenological quantities $T_1$, $T_2$ and do not focus on the sources of decoherence in the system.

Probably the first experiments that revealed inadequacy of Bloch equations in transient regime were ESR measurements of $E_1'$-centers in glassy silica [16,17] and of $[AlO_4]^0$ centers in quartz [16]. When $\sigma \gg \Omega_R$, the detected RO were of the form $M_\perp(t) \sim e^{-t/\tau_R} J_0(\Omega_R t)$, and the decay rates contained two terms,

$$\tau_R^{-1} = \alpha + \beta\Omega_R, \quad (18)$$

with constant dimensionless coefficient $\beta \sim 10^{-2}$.

Later, the decay rates of the form (18) were obtained for a variety of electronic spin systems, including rare earths [5,10,18] and transition metal ions [19,20].

In view of all above-mentioned, we address the 2nd term in Eq. (18) to relaxation induced by MD interactions between the spins of the same resonant line coherently driven by MW field (see Eq. (14)), and the 1st term (usually indistinguishable) to other spin relaxation processes that do not depend on $\Omega_R$ (e.g. spin-lattice interactions). Up to the moment, the only explanation of the term $\beta\Omega_R$ was given in the framework of semi-phenomenological model of Shakhmuratov et al [21]. They suggested that the driving MW field $B_1 e^{-i\omega_0 t}$ induces stochastic local magnetic field of the amplitude proportional to $B_1$ and included the term $\sim B_1$ into Bloch equations. Since the measured parameter $\beta$ was concentration-dependent, they ascribed the stochastic field to MD interactions between paramagnetic ions [17]. However, they have not given any estimation of the amplitude of the stochastic field.

In order to reveal the importance of the above decay mechanism in transient response of real systems, we have calculated RO decay rates $\tau_R^{-1} = \beta_d \Omega_R$ for $E_1'$ and $[AlO_4]^0$ centers and compared them to the ones observed in [16,17]. The parameter $\beta_d$ was obtained from the average slope angle $\psi$ with respect to $\Omega_R$ axis of the calculated $\tau_R^{-1}(\Omega_R)$ dependence (Fig. 2) in the experimental range of $\Omega_R$: $\text{tg}\psi(E_1') = 0.5$, $\text{tg}\psi(AlO_4) = 0.95$. Then $\beta_d = \kappa\Delta\omega_d \text{tg}\psi/\sigma$, where $\kappa$ was the relative concentration of spins belonging to the given resonance line, $\kappa\left(E_1'\right)=1$, $\kappa\left(\mathrm{AlO}_4\right)=1/6$ (in the latter case there were 6 well-resolved hyperfine lines). ESR lines were approximated by Gaussian distribution with experimental standard deviations $\sigma\left(E_1'\right)=2\pi\cdot 1\,\mathrm{MHz}$, $\sigma\left(\mathrm{AlO}_4\right)=2\pi\cdot 0.25\,\mathrm{MHz}$ [16,17]. Since $\beta_d \sim \Delta\omega_d \sim C$, we write $\beta_d = \gamma_d C$, with $\gamma_d = \dfrac{4\pi^2 g_e^2 \mu_B^2 \kappa \,\mathrm{tg}\,\psi}{9\sqrt{3}\hbar\sigma}$; then $\gamma_d\left(E_1'\right)=1.25\cdot 10^{-19}\,\mathrm{cm}^3$ and $\gamma_d\left(\mathrm{AlO}_4\right)=4.3\cdot 10^{-20}\,\mathrm{cm}^3$. The measured RO decay parameters $\beta$ are found to be well approximated within experimental error bounds by the sum $\beta = \beta_0 + \beta_d$ (see Table 1), with $\beta_0\left(E_1'\right)=5\cdot 10^{-2}$ and $\beta_0\left(\mathrm{AlO}_4\right)=2.11\cdot 10^{-2}$. Here the term $\beta_d$ linear in spin concentration is induced by MD interactions, and the remaining constant term $\beta_0$ originates from the contributions of other kind (probably MW field inhomogeneities inside the sample volume [10]). Our model predicts that for spin concentrations $C > 5\cdot 10^{17}\,\mathrm{cm}^{-3}$ in both systems $\beta_d$ will prevail over $\beta_0$. The calculated values of $\beta_d$ agree very well with the concentration-dependent parts of $\beta$ (cf. columns 4 and 6 in Table 1), which experimentally validates the RO decay mechanism considered in the present paper.

## IV. CONCLUSIONS

We performed first-principles calculations of the relaxation processes occurring in dipolar-coupled *A* spin bath during transient regime. We have shown that the decoherence in *A* spin bath, under specific conditions, is enhanced by the external resonant microwave field, in contrast with *B* spin bath. The decay rate of Rabi oscillations depends linearly on the concentration of magnetic impurity and, in the strong inhomogeneity case, almost linearly on Rabi frequency. The predictions of the model are in quantitative agreement with concentration-dependent parts of experimentally observed decay rates [16,17]. The obtained results are of importance for the implementation of many-qubit systems in quantum information processing.

## ACKNOWLEDGMENTS

This work was supported by RFBR (grant N09-02-00930) and by Dynasty Foundation. Author is grateful to B. Z. Malkin for advice and fruitful discussions of the results.

**References**

[1] I. I. Rabi, Phys. Rev. **51**, 652 (1937).

[2] H. C. Torrey, Phys. Rev. **76**, 1059 (1949).

[3] P. W. Atkins, A. J. Dobbs and K. A. McLauchlan, Chem. Phys. Lett. **25**, 105 (1974).

[4] G. B. Hocker and C. L. Tang, Phys. Rev. Lett. **21**, 591 (1968).

[5] R. M. Rakhmatullin, I. N. Kurkin, G. V. Mamin, S. B. Orlinskii, M. R. Gafurov, E. I. Baibekov, B. Z. Malkin, S. Gambarelli, S. Bertaina, and B. Barbara, Phys. Rev. B **79**, 172408 (2009).

[6] S. Bertaina, S. Gambarelli, T. Mitra, B. Tsukerblat, A. Müller, and B. Barbara, Nature **453**, 203 (2008).

[7] D. P. DiVincenzo, Science **270**, 255 (1995).

[8] J. I. Cirac and P. Zoller, Phys. Rev. Lett. **74**, 4091 (1995).

[9] J. J. L. Morton, A. M. Tyryshkin, A. Ardavan, K. Porfyrakis, S. A. Lyon, and G. A. D. Briggs, Phys. Rev. A **71**, 012332 (2005).

[10] E.I. Baibekov, I.N. Kurkin, M.R. Gafurov, B. Endeward, R.M. Rakhmatullin, G.V. Mamin, to be published.

[11] A. Abragam, *The Principles of Nuclear Magnetism*, Clarendon Press, Oxford (1961).

[12] B. Herzog and E. L. Hahn, Phys. Rev. **103**, 148 (1956).

[13] W.B. Mims, Phys. Rev. **168**, 370 (1968).

[14] V. V. Dobrovitski, A. E. Feiguin, R. Hanson, and D. D. Awschalom, Phys. Rev. Lett. **102**, 237601 (2009).

[15] A. Abragam and B. Bleaney, *Electron Paramagnetic Resonance of Transition Ions*, Oxford Univ. Press, London (1970).

[16] R. Boscaino, F. M. Gelardi, and J. P. Korb, Phys. Rev. B **48**, 7077 (1993).

[17] S. Agnello, R. Boscaino, M. Cannas, F. M. Gelardi, and R. N. Shakhmuratov, Phys. Rev. A **59**, 4087 (1999).

[18] S. Bertaina, S. Gambarelli, A. Tkachuk, I. N. Kurkin, B. Malkin, A. Stepanov, and B. Barbara, Nature Nanotech. **2**, 39 (2007).

[19] S. Nellutla, K.-Y. Choi, M. Pati, J. van Tol, I. Chiorescu, and N. S. Dalal, Phys. Rev. Lett. **99**, 137601 (2007).

[20] J. H. Shim, S. Gambarelli, S. Bertaina, T. Mitra, B. Tsukerblat, A. Müller, and B. Barbara, arXiv:1006.4960v2.

[21] R. N. Shakhmuratov, F. M. Gelardi, and M. Cannas, Phys. Rev. Lett. **79**, 2963 (1997).

Table 1. Measured ($\beta$, [16,17]) and calculated ($\gamma_d C$, this work) RO decay parameters for $E_1'$ centers in glassy silica and $[AlO_4]^0$ centers in quartz; $\gamma_d(E_1') = 1.25 \cdot 10^{-19} \text{cm}^3$, $\gamma_d(AlO_4) = 4.3 \cdot 10^{-20} \text{cm}^3$. The last column represents the best fit of $\beta$ within the concentration error bounds by the function $\beta = \beta_0 + \gamma_d C$, with $\beta_0(E_1') = 5 \cdot 10^{-2}$ and $\beta_0(AlO_4) = 2.11 \cdot 10^{-2}$.

| Spin centers | Sample no. | $C, 10^{16} \text{cm}^{-3}$ | $\beta, 10^{-2}$ | $\gamma_d C, 10^{-2}$ | $\beta_0 + \gamma_d C, 10^{-2}$ |
|---|---|---|---|---|---|
| $E_1'$ | 1 | $7.5 \pm 2$ | $4.8 \pm 0.5$ | $0.94 \pm 0.25$ | 5.7 |
|  | 2 | $16 \pm 5$ | $6.1 \pm 0.5$ | $2.00 \pm 0.63$ | 6.4 |
|  | 3 | $24 \pm 8$ | $10.6 \pm 0.5$ | $3.00 \pm 1.00$ | 9.0 |
| $[AlO_4]^0$ | 1 | $4.0 \pm 0.4$ | $2.4 \pm 0.1$ | $0.17 \pm 0.02$ | 2.30 |
|  | 2 | $1.0 \pm 0.1$ | $2.1 \pm 0.1$ | $0.043 \pm 0.004$ | 2.15 |

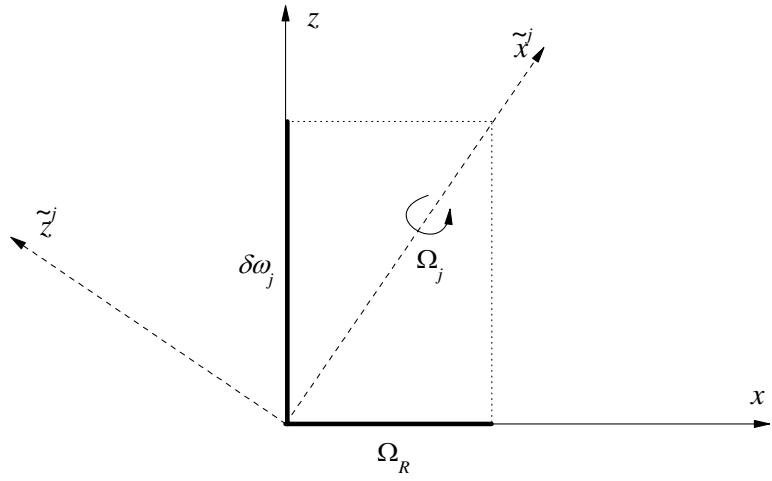

FIG. 1. A transformation of the local basis $x^j, y^j, z^j \to \tilde{x}^j, \tilde{y}^j, \tilde{z}^j$ for $j$-th spin detuned by $\delta\omega_j$ from resonance.

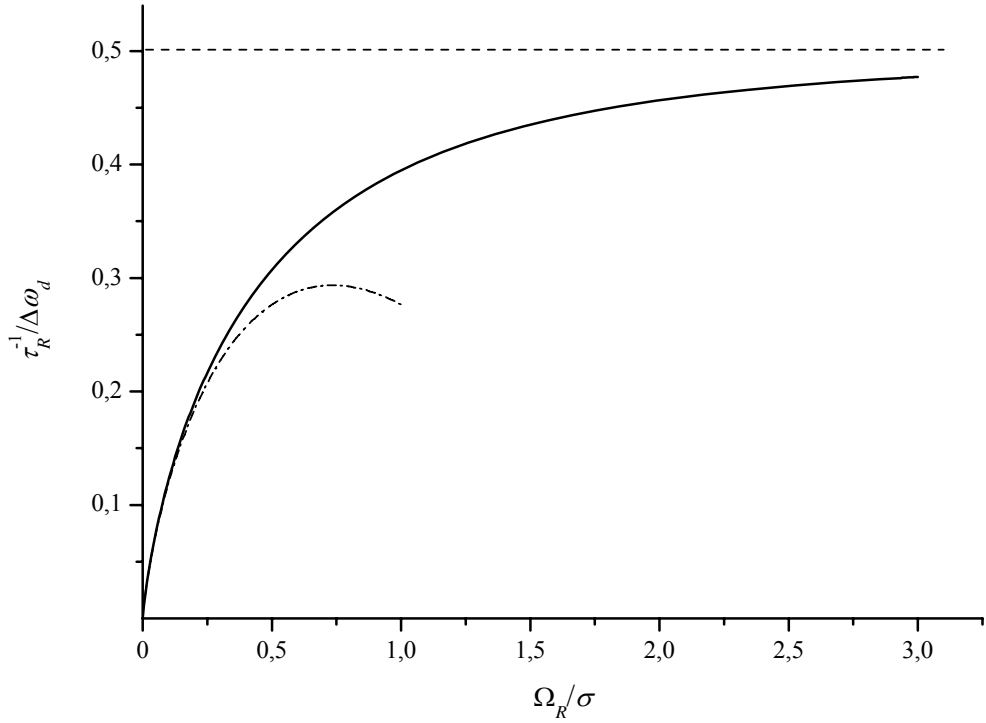

FIG. 2. The dependence of decay rate $\tau_R^{-1}$ of transverse magnetization on Rabi angular frequency $\Omega_R$ calculated from Eq. (12) assuming Gaussian line shape (solid line). The asymptotes $\tau_R^{-1} = \Delta\omega_d/2$ for $\sigma \ll \Omega_R$ and $\tau_R^{-1} = \Delta\omega_d g_G(\omega_0)\Omega_R \ln\dfrac{2\sigma}{\Omega_R}$ for $\sigma \gg \Omega_R$ are represented by dashed and dash-dotted lines, respectively. For $\sigma \gg \Omega_R$, the function $\tau_R^{-1}(\Omega_R)$ is nearly linear in the specific range of $\Omega_R$.

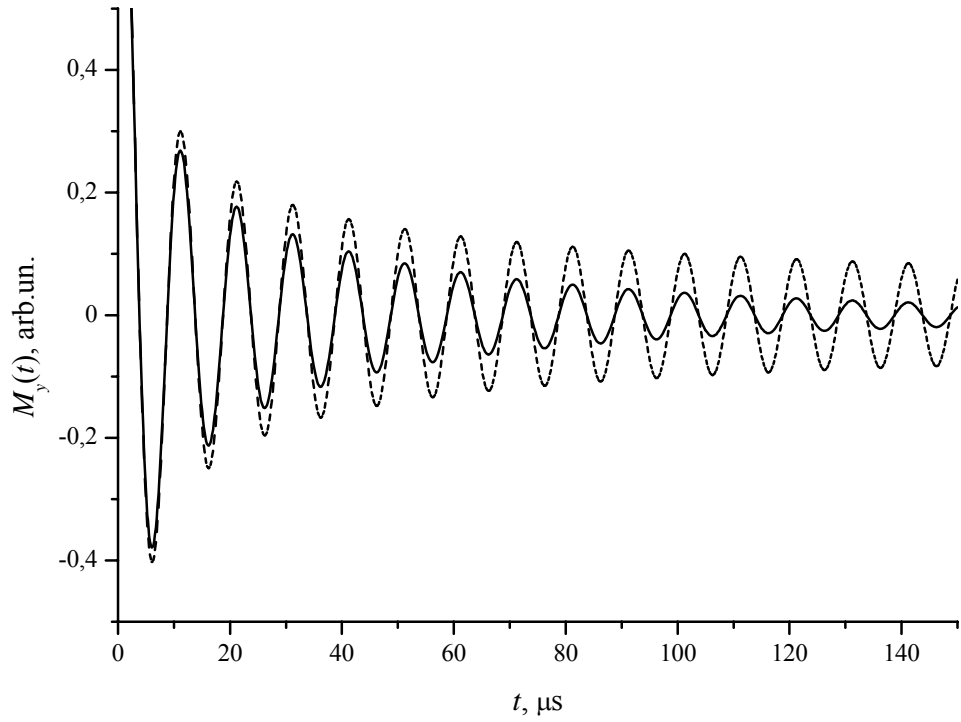

FIG. 3. The time dependence of transverse magnetization as calculated from Eq. (14), with spin concentration $C = 10^{17} \text{cm}^{-3}$, standard deviation of Gaussian distribution $\sigma/2\pi = 1\,\text{MHz}$ and Rabi frequency $\Omega_R/2\pi = 100\,\text{kHz}$ (solid line). Dashed line represents the dependence $J_0(\Omega_R t)$ that would take place in the absence of magnetic dipole interactions.